\documentclass[11pt,fleqn]{article}
\usepackage{graphicx}
\usepackage{latexsym}
\title{Maximal complexity of finite words}
\author{Mira-Cristiana Anisiu\footnote{Tiberiu Popoviciu Institute 
of Numerical Analysis, Romanian Academy, Cluj-Napoca, 
E-mail: \texttt{mira@math.ubbcluj.ro}}  \and 
Zolt\'an Bl\'azsik\footnote{Bolyai Institute of Mathematics, 
University of Szeged, E-mail: \texttt{blazsik@sol.cc.u-szeged.hu}}
\and
Zolt\'an  K\'asa\footnote{Faculty of Mathematics and In\-for\-matics, Babe\c s-Bolyai University of Cluj-Napoca, E-mail: \texttt{kasa@cs.ubbcluj.ro}}}

\newtheorem{thm}{Proposition}
\newtheorem{lem}{Lemma}
\newtheorem{rem}{Remark}

\begin{document}
\maketitle

\abstract{
The \emph{subword complexity} of a finite word $w$ of length $N$ is 
a function which associates to each $n\le N$ the number 
of all distinct subwords of $w$ having the length $n$.
We define the \emph{maximal complexity} $C(w)$ as the maximum of the subword
complexity for $n \in \{1,2,\ldots ,N \}$, and the  
\emph{global maximal complexity} $K(N)$ as the maximum of $C(w)$ for all 
words $w$ of a fixed length $N$ over a finite alphabet. By
$R(N)$ we will denote the set of the values $i$ for which there exits a word 
of length $N$ having $K(N)$ subwords of length $i$.
 $M(N)$ represents the number of words of length $N$ whose maximal complexity is 
equal to the global maximal complexity.
 
The values of $K(N)$ and $R(N)$ are obtained; methods to compute $M(N)$ using the de Bruijn graphs and trees are given. An open problem is to find a formula for $M(N)$.}

\section{Introduction}
A \emph{finite word} is a finite sequence of letters over a finite alphabet 
${\cal A}$, and can be represented as a concatenation of its letters:

\[w=w_1w_2\ldots w_N \quad \textrm{ with } w_i\in {\cal A} \textrm{ for } 
1\le i\le N.
\]

The number $N$ is the length of $w$  and is denoted by $|w|$. A word with no 
letters (i.e. of length 0) is the \emph{empty word}, denoted by $\varepsilon$.
We denote by  ${\cal A}^+$ the set of nonempty words over ${\cal A}$, by 
${\cal A}^*= {\cal A}^+ \cup \{\varepsilon\}$ the set of words over ${\cal A}$
and by ${\cal A}^n$ the set of words of length $n$ over ${\cal A}$.

A word $u$ is a \emph{factor} (or \emph{subword}) of $w$ if there exist words
$x,y \in {\cal A}^*$ such that $w=xuy$. If $x\ne \varepsilon$ and 
$y\ne \varepsilon$ then $u$ is a \emph{proper factor} (\emph{proper subword}) 
of $w$.  If $x=\varepsilon$ ($y=\varepsilon$) then $u$ is a \emph{prefix} 
(\emph{suffix}) of $w$. Let us denote by $F(w)$ the set of all nonempty factors
of $w$, and by $F_n(w)$ the set of all factors of $w$ of length $n$ 
(hence $F_n(w)=F(w) \cap {\cal A}^n$). 

The \emph{subword complexity} of $w$ counts the number of all distinct factors
of a given length occurring in $w$ and is defined as

\[ f_w(n)=\textrm{Card}(F_n(w)) \qquad \textrm{ for } 1\le n \le |w|.
\]
Clearly $f_w(1)\le \textrm{Card}({\cal A})$ and we can consider $f_w(n)=0$ for $n>|w|$. The 
subword complexity has been extensively studied in \cite{hei}, \cite{lothaire} and
 \cite{aldo}.

The maximal value of the subword complexity $f_w(n)$ for $1\le n \le |w|$ 
is called the  \emph{maximal complexity} of $w$ and is denoted by $C(w)$:

\[ C(w)=\max \{f_w(n) \; | \; n\ge 1 \}.
\]  

The \emph{global maximal complexity} in ${\cal A}^N$ is equal to

\[ K(N)= \max \{ C(w) \; | \; w\in {\cal A}^N \}.
\]

We shall denote by $R(N)$ the set of values $i$ for which there exists a word
$w\in {\cal A}^N$ such that $f_w(i)=K(N)$:

\[  R(N)= \{i\in \{1,2,\ldots, N\} \;|\; \exists w\in {\cal A}^N : 
    f_w(i)= K(N) \}.
\]

The number of words in ${\cal A}^N$ with the maximal complexity equal to the 
global maximal complexity will be denoted by $M(N)$:

\[ M(N)= \textrm{Card}(\{w\in {\cal A}^N : C(w)=K(N)\}).
\]

\begin{rem}\hspace*{-0.60em}\textbf{.} \label{rem1}
\emph{
If $\textrm{Card}({\cal A})=q$, for $q=1$ the only word of length $N$ is
$w_0=\underbrace{0\ldots 0}_N$ for which $f_{w_0}(i)=1,$  
$i\in \{1,2,\ldots, N\}$, hence $C(w)=1=K(N)$, $R(N)=\{1,2,\ldots ,N\}$ and 
$M(N)=1.$ For $q\ge 2$, but $N\le q$, for each word $w_1$ which contains $N$ 
distinct elements of ${\cal A}$ we have $C(w_1)=f_{w_1}(1)=N=K(N)$,
$R(N)=\{1\}$ and $M(N)=P^N_q$ (permutations of $N$ elements taken from $q$).}
\end{rem}

Some values for $K(N)$, $R(N)$ and $M(N)$ in the case of an alphabet of 2 
letters are given in Table \ref{table1}. In the case $N=3$ the 
following six words have maximal complexity: 001, 010, 011, 100, 101, 110.
For each of them $f_w(1)=2, f_w(2)=2, f_w(3)=1$, so $K(3)=2, R(3)=\{1,2\}$ and 
$M(3)=6$.

\begin{figure}[!ht]
\begin{center}
\begin{tabular}{|r|r|r|r|}\hline
$\quad N\quad$ & $\quad K(N)\quad$ & $\quad R(N)\quad$ & 
   $\quad M(N)\quad$ \\ \hline\hline
1  & 1  & 1    & 2 \\
2  & 2  & 1    & 2 \\ \hline
3  & 2  & 1, 2 & 6 \\
4  & 3  & 2    & 8 \\
5  & 4  & 2    & 4 \\ \hline
6  & 4  & 2, 3 & 36 \\
7  & 5  & 3    & 42 \\
8  & 6  & 3    & 48 \\
9  & 7  & 3    & 40 \\
10 & 8  & 3    & 16 \\ \hline
11 & 8  & 3, 4  & 558 \\
12 & 9   & 4     & 718 \\
13 & 10  & 4     & 854 \\ 
14 & 11  & 4     &  920\\
15 & 12  & 4     &  956\\
16 & 13  & 4     &  960\\
17 & 14  & 4     &  912\\
18 & 15  & 4     &  704\\
19 & 16  & 4     &  256\\ \hline
20 & 16  & 4, 5  &  79006\\
\hline\hline
\end{tabular}
\end{center}
\label{table1}
\centerline{Table \ref{table1}}
\end{figure}

\section{Global maximal subword complexity of finite words}
In this section we shall compute the values of the global maximal complexity
$K(N)$, as well as those of $R(N)$, proving that they are in agreement with
the values in Table 1. Some special cases being solved in Remark \ref{rem1},
in what follows we shall consider alphabets with 
$\textrm{Card}({\cal A})=q\ge 2$ and 
words of length $N>q$.

We shall use the following result.

\begin{lem}\hspace*{-0.60em}\textbf{.} \emph{\cite{martin}} \label{lem1}
For each $k\in N^*$, the shortest word containing all the $q^k$ words of 
length $k$ has $q^k+k-1$ letters (hence in this word each of the $q^k$ 
words of length $k$ appears only once).
\end{lem}

An algorithm for obtaining such a word for ${\cal A}= \{e_1,e_2,\ldots, e_q\}$
is the following \cite{martin}:

\medskip
\emph{i.} Each of the first $k-1$ symbols is equal to $e_1$. 

\emph{ii.} If the sequence $a_1a_2\ldots a_k\ldots a_{m-k+1}\ldots a_{m-1}$
(with $a_1=\ldots =a_{k-1}=e_1$, $m\ge k$ and the $a$'s representing 
the $e$'s in a certain order) has been obtained, the symbol $a_m$ to be
added is the $e_i$ with the greatest subscript possible such that 
$a_{m-k+1}\ldots a_{m-1}a_m$ does not duplicate a previously occurring 
section of $k$ symbols in the above sequence.

\emph{iii.} Rule \emph{ii} is first applied for $m=k$ (in which case 
$a_m=a_k=e_q$) and then applied repeatedly until a further application 
is impossible.

\medskip
\begin{thm}\hspace*{-0.60em}\textbf{.} \label{thm1}
If $\emph{Card}({\cal A})=q$ and  $q^k+k \le N\le q^{k+1}+k$ then $K(N)= N-k$.
\end{thm}

\noindent\textbf{Proof.}
Let us consider at first the case $N=q^{k+1}+k$, $k\ge 1$.
 
From Lemma \ref{lem1} we obtain the existence of
a word $W$ of length $q^{k+1}+k$ which contains all the $q^{k+1}$ words
of length $k+1$, hence $f_W(k+1)=q^{k+1}$. It is obvious that 
$f_W(l)=q^l<f_W(k+1)$ for $l\in \{1,2,\ldots, k\}$ and 
$f_W(k+1+j)=q^{k+1}-j< f_W(k+1)$ for $j\in \{1,2,\ldots q^{k+1}-1 \}$.
 Any other word of length $q^{k+1}+k$ will have the maximal complexity 
less than or equal to $C(W)=f_W(k+1)$, hence we have $K(N)=q^{k+1}=N-k$.

For $k\ge 1$ we consider now the values of $N$ of the form 
$N=q^{k+1}+k-r$ with $r\in \{1,2,\ldots, q^{k+1}-q^k\}$, hence 
$q^k+k\le N< q^{k+1}+k$. If from the word $W$ of length $q^{k+1}+k$
considered above we delete the last $r$ letters, we obtain a word $W_N$ 
of length $N=q^{k+1}+k-r$ with $r\in \{1,2,\ldots, q^{k+1}-q^k\}$. 
This word will have $f_{W_N}(k+1)=q^{k+1}-r$ and this value will be 
its maximal complexity. Indeed, it is obvious that $f_{W_N}(k+1+j)=
      f_{W_N}(k+1)-j < f_{W_N}(k+1)$ for $j\in \{1,2,\ldots, N-k-1\}$;
for $l\in \{1,2,\ldots , k\}$ it follows that
$f_{W_N}(l)\le q^l\le q^k\le q^{k+1}-r= f_{W_N}(k+1)$, hence 
$C(W_N)=f_{W_N}(k+1)=q^{k+1}-r$. Because it is not possible for a 
word of length $N=q^{k+1}+k-r$, with $r\in \{1,2,\ldots, q^{k+1}-q^k\}$
to have the maximal complexity greater than $q^{k+1}-r$, it follows that
$K(N)= q^{k+1}-r=N-k$.

\begin{thm}\hspace*{-0.60em}\textbf{.} \label{thm2}
If $\emph{Card}({\cal A})=q$ and  $q^k+k < N < q^{k+1}+k+1$ then $R(N)= \{k+1\}$;
if $N=q^k+k$ then $R(N)= \{k,k+1\}$.
\end{thm}

\noindent\textbf{Proof.} In the first part of the proof of Proposition 
\ref{thm1}, we proved for $N=q^{k+1}+k$, $k\ge 1$, the existence of a word $W$ of 
length $N$ for which $K(N)=f_W(k+1)=N-k$. This means that $k+1\in R(N)$.
For the word $W$, as well as for any other word $w$ of length $N$, we have 
$f_w (l) < f_W(k+1)$, $l\not = k+1$, because of the special construction 
of $W$, which  contains all the words of length $k+1$ in the most compact way.
It follows that $R(N)=\{k+1\}$.

As in the second part of the proof of Proposition 1, we consider 
$N=q^{k+1}+k-r$ with $r\in \{1,2,\ldots q^{k+1}-q^k\}$ and the word $W_N$ for 
which $K(N)=f_{W_N}(k+1)=q^{k+1}-r$. We have again $k+1\in R(N)$. For 
$l>k+1$, it is obvious that the complexity function of $W_{N}$, or of any 
other word of length $N$, is strictly less than $f_{W_{N}}(k+1)$. We examine
now the possibility of finding a word $W$ with $f_W(k+1)=N-k$ for which
$f_W(l)=N-k$ for $l\le k$. We have $f_W(l)\le q^l\le q^k\le q^{k+1}-r$, hence 
the equality $f_W(l)=N-k=q^{k+1}-r$ holds only for $l=k$ and $r=q^{k+1}-q^k$, 
that is for $N=q^k+k$. 
We show that for $N=q^k+k$
we have indeed $R(N)=\{k,k+1\}$. If we start with Martin's word of length
$q^k+k-1$ (or with another de Bruijn word) and add to this any letter from 
${\cal A}$, we obtain obviously a word $V$ of length $N=q^k+k$, which 
contains all the $q^k$ words of length $k$ and $q^k=N-k$ words of length 
$k+1$, hence $f_V(k)=f_V(k+1)=K(N).$

\begin{rem}\hspace*{-0.60em}\textbf{.} 
\emph{ Having in mind the algorithm given by Martin \cite{martin} 
(or other more efficient algorithms), words $w$ 
with maximal complexity $C(w)=K(N)$ can be easily constructed for each
$N$ and for both situations in Proposition \ref{thm2}.
}
\end{rem}

\section{De Bruijn graphs and trees}

In the previous section the global maximal complexity $K(N)$ for words
of length $N$ was obtained, as well as the set of points $R(N)$ where 
$K(N)$ is equal to the maximal value of the subword complexity of certain
words of length $N$. To this aim we used a special word constructed by Martin 
\cite{martin}, which is one of the de Bruijn words. 
A de Bruijn word for given $q$ and $k$
is  a word over an alphabet with $q$ letters, containing all 
$k$-length words exactly once. The length of such a word is $q^k+k-1$. 

In order to tackle the 
problem of finding the number of the words for which the global maximal
complexity is attained, we shall use the de Bruijn graphs and trees.
 
\begin{figure}[!ht]
\begin{center}
\includegraphics[scale=0.8]{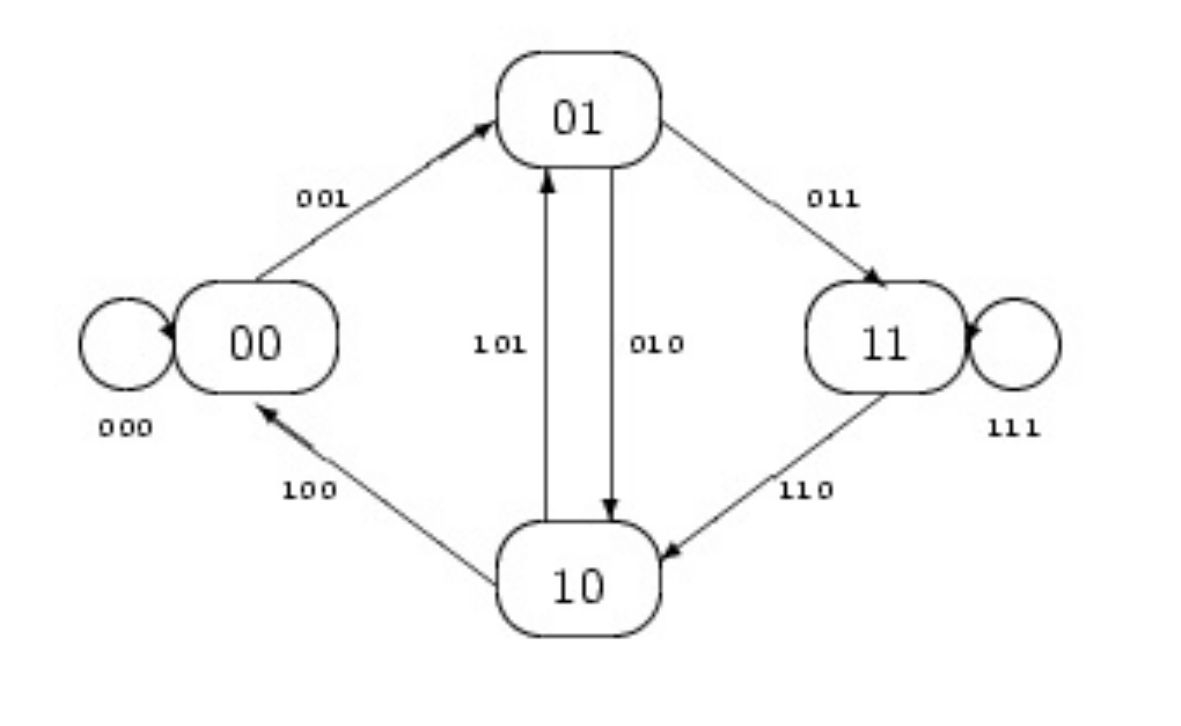}
\end{center}
\caption{The de Bruijn graph $B(2,2)$.\label{fig1}}
\end{figure}

\begin{figure}[!ht]
\begin{center}
\includegraphics[scale=0.5]{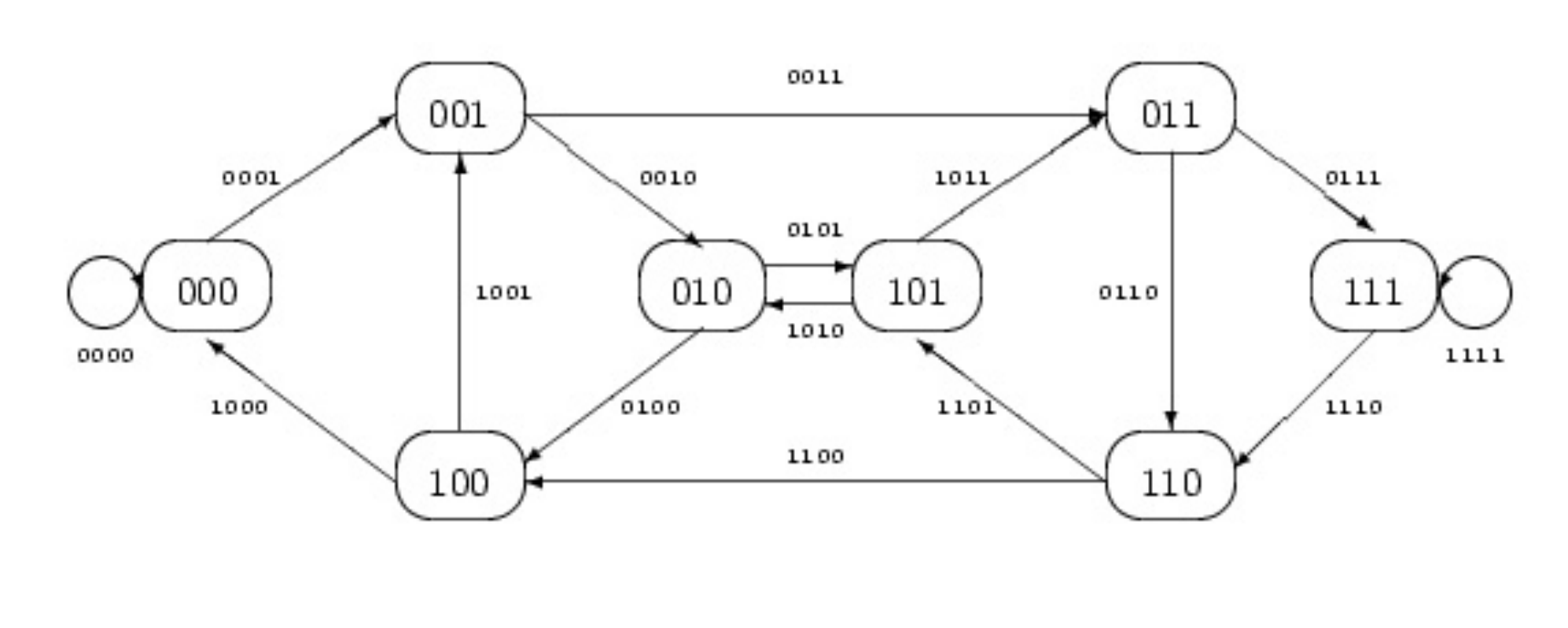}
\end{center}
\caption{The de Bruijn graph $B(2,3)$.\label{fig2}}
\end{figure}

For a $q$-letter alphabet ${\cal A}$ the de Bruijn graph is defined as:
\[ 
B(q,k)= (V(q,k), E(q,k))
\]
with $V(q,k)={\cal A}^k$ as the set of vertices,
and $E(q,k)= {\cal A}^{k+1}$ as the set of directed arcs. There is
an arc from $x_1x_2\ldots x_k$ to $y_1y_2\ldots y_k$ if 
$x_2x_3\ldots x_k=y_1y_2\ldots y_{k-1}$, and this arc is denoted by
$x_1x_2\ldots x_ky_k$. See Fig. \ref{fig1} and \ref{fig2} for $B(2,2)$ and 
$B(2,3)$. The de Bruijn graphs $B(q,k)$ are nonplanar for $k\ge 4$, $q\ge 2$.

In the de Bruijn graph $B(q,k)$ a path (i. e. a walk with distinct vertices)
\[ a_1a_2\ldots a_k, \quad a_2a_3\ldots a_{k+1},\; \ldots ,\; 
a_{r-k+1}a_{r-k+2}\ldots a_r  \qquad (r>k) \] 
corresponds to an $r$-length word $a_1a_2\ldots a_ka_{k+1}\ldots a_r$, which 
is obtained by adding, in turn, to the vertex $a_1a_2\ldots a_k$ the last
letter of the following vertices in the path. 
For example in $B(2,3)$ the path $001$, $010$, $101$ corresponds to the 
word $00101$.  
Every maximal length path in the graph $B(q,k)$ (which is a
Hamiltonian one) corresponds to a de Bruijn word.

In the directed graph $B(q,k)$ there always exists an Eulerian circuit because
it is connected and  all its
vertices have the same indegree and outdegree $q$. An Eulerian
circuit in $B(q,k)$  is a Hamiltonian path in $B(q,k+1)$ (which always can
be continued in a Hamiltonian cycle). For example in $B(2,2)$ the following
walk: $000$, $001$, $010$, $101$, $011$, $111$, $110$, $100$ represents an
Eulerian circuit, which in $B(2,3)$ is a Hamiltonian path.

In order to study the number of words in ${\cal A}^k$ which have the maximal
complexity equal to the global maximal complexity $K(N)$ we shall introduce 
the so-called de Bruijn trees. 
A \emph{de Bruijn tree} $T(q,w)$ with the root $w\in {\cal A}^k$ is a $q$-ary tree defined 
recursively as follows:

\medskip
\emph{i.} The $k$-length word $w$ over the alphabet 
${\cal A}= \{e_1,e_2,\ldots. e_q\}$ 
is the root of $T(q,w)$.

\emph{ii.} If at any step of the recursive construction of the tree,
$x_1x_2\ldots x_k$ is a (temporary) leaf (a vertex with outdegree equal to 0),
then each word among $x_2x_3\ldots x_ke_1$, 
$x_2x_3\ldots x_ke_2$, \ldots , $x_2x_3\ldots x_ke_q$ which is not in the path 
from the root to $x_1x_2\ldots x_k$ will be a descendant of $x_1x_2\ldots x_k$. 
 
\emph{iii.} The rule \emph{ii} is applied as many times as it is possible.

\begin{figure}
\begin{center}
\includegraphics[scale=0.5]{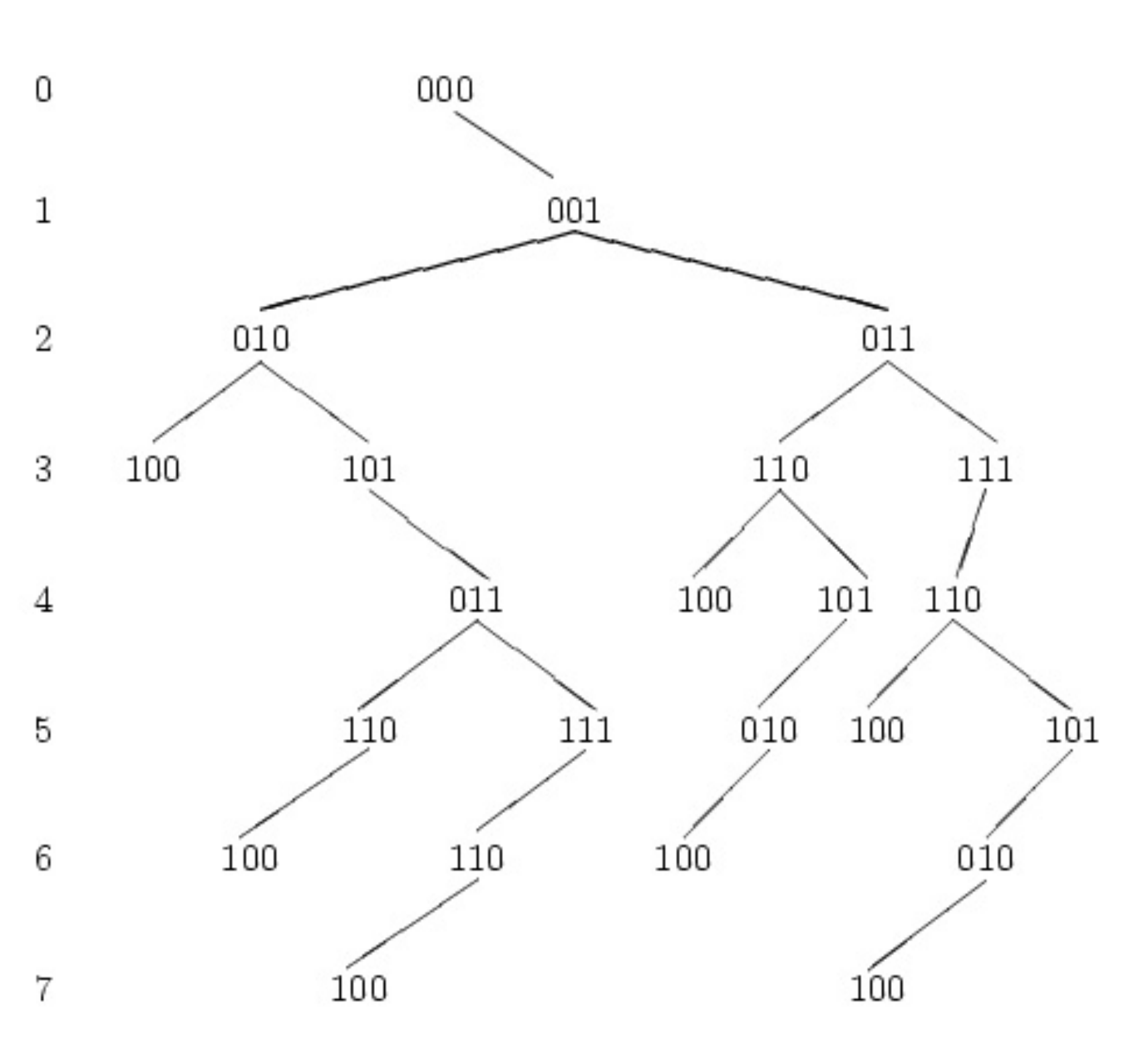}
\end{center}
\caption{De Bruijn tree $T(2,000)$.}
\label{fig3a}
\end{figure}

\begin{figure}
\begin{center}
\includegraphics[scale=0.5]{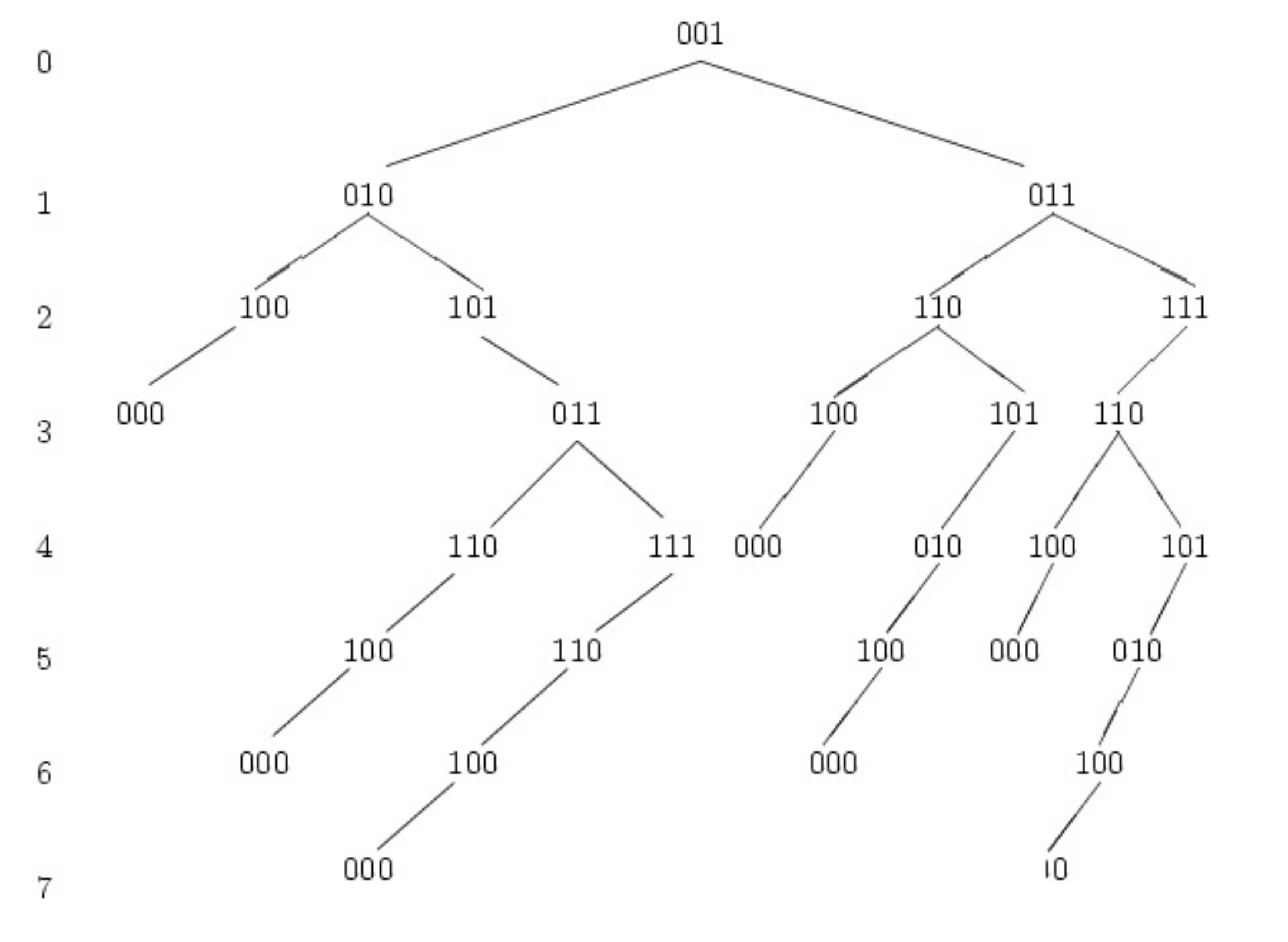}
\end{center}
\caption{De Bruijn tree $T(2,001)$.}
\label{fig3b}
\end{figure}

\begin{figure}
\begin{center}
\includegraphics[scale=0.5]{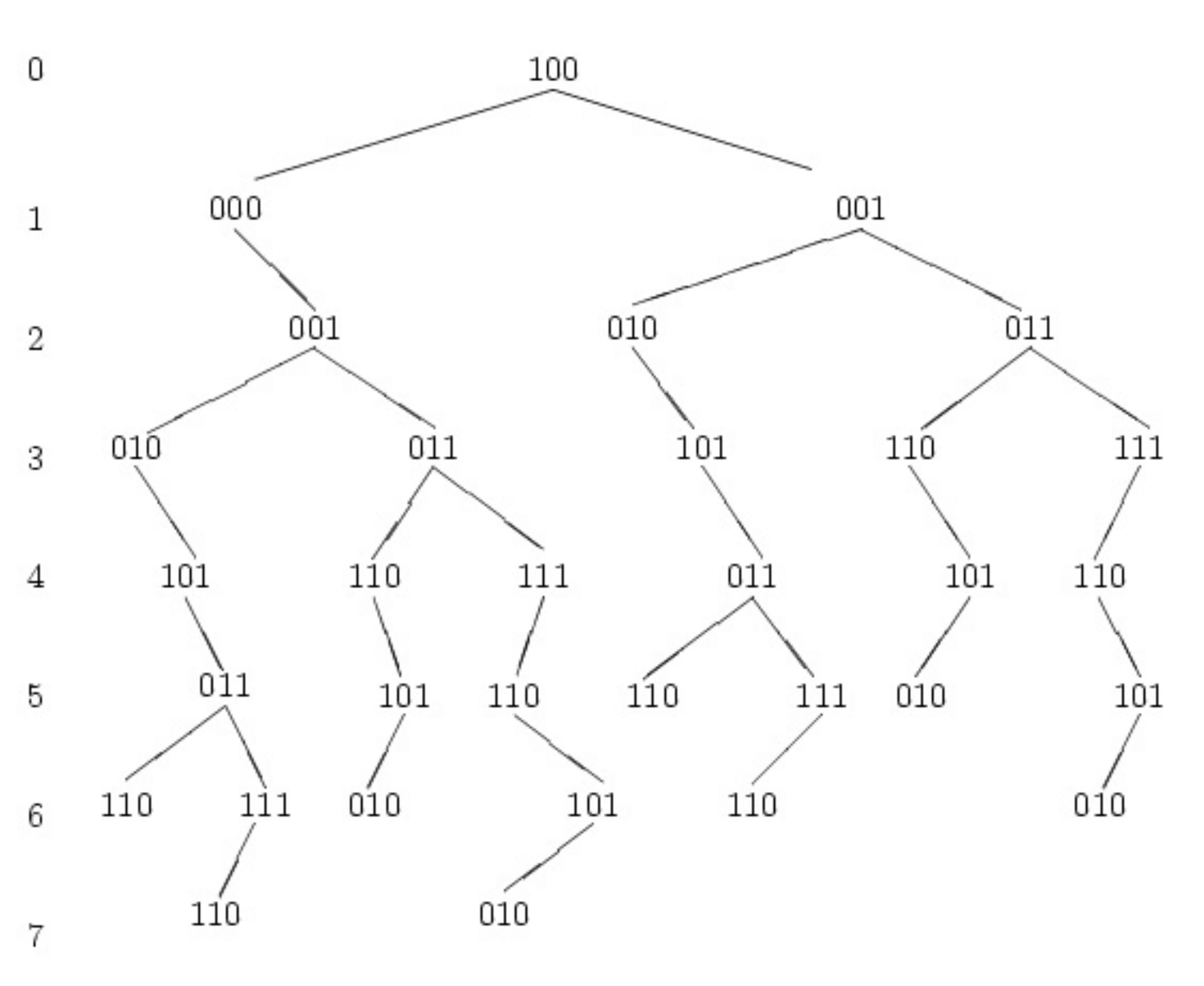}
\end{center}
\caption{De Bruijn tree $T(2,100)$.}
\label{fig3c}\end{figure}

\begin{figure}
\begin{center}
\includegraphics[scale=0.5]{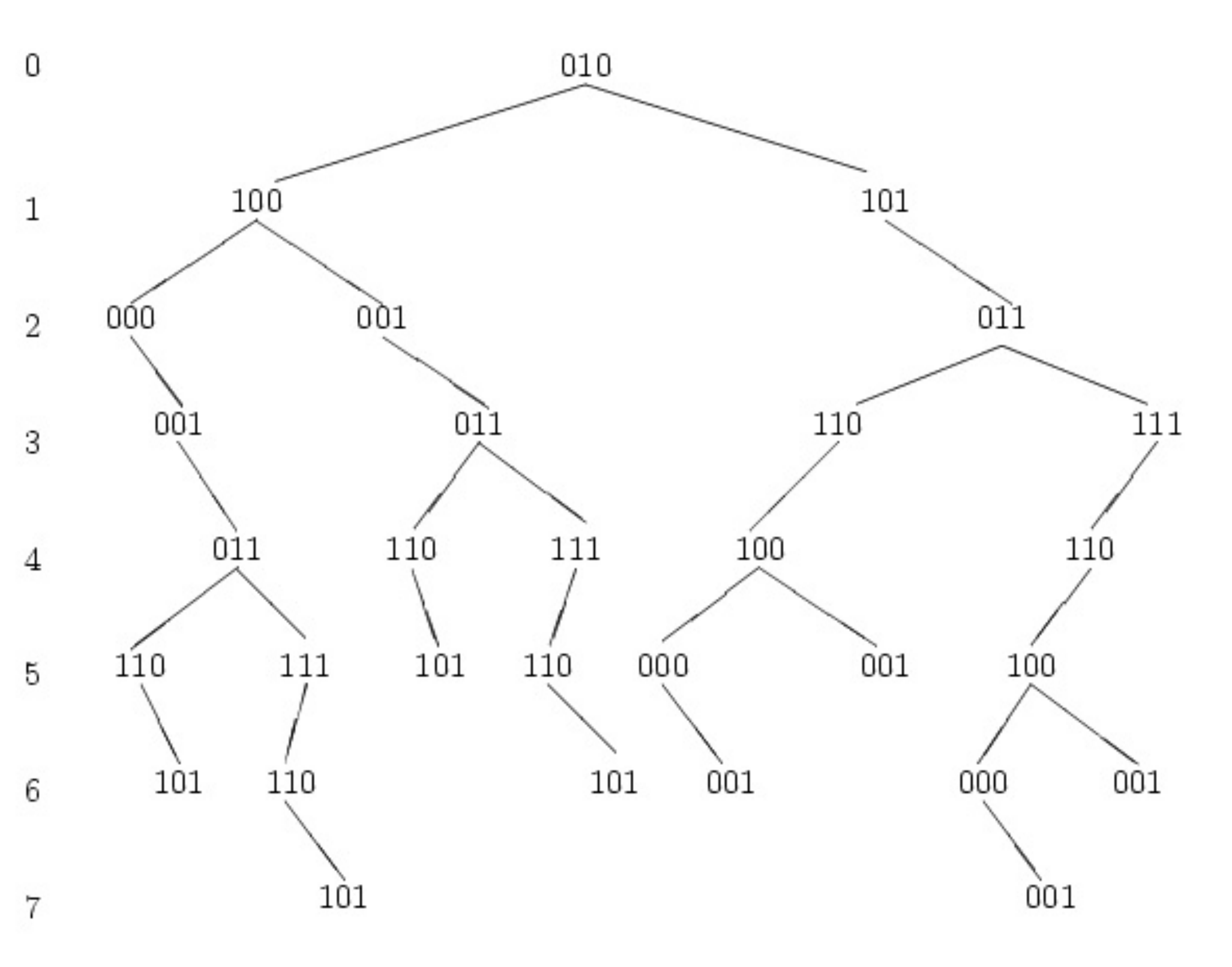}
\end{center}
\caption{De Bruijn tree $T(2,010)$.}
\label{fig3d}
\end{figure}

\medskip
A path is maximal if we cannot add an arc to its beginning or to its end
without destroying  the path property. If a maximal path is of maximal length 
then it is a Hamiltonian one. In any de Bruijn tree each branch is a 
maximal path in the de Bruijn graph $B(q,k)$ which begins with the root, 
and all maximal paths beginning with the root occur.  
For the de Bruijn trees $T(2,000)$, $T(2,001)$, $T(2,010)$ and $T(2,100)$ 
 see Fig. \ref{fig3a}--\ref{fig3d}. The word obtained by Martin's algorithm corresponds to
the branch of maximal length in the right side of the de Bruijn tree 
$T(2,001).$

\medskip

\section{Methods to compute $M(N)$}
The number $M(N)$ of the words of length $N$ for which the maximal complexity is equal to the global maximal complexity $K(N)$ can be expressed both in terms of certain paths in a de Bruijn graph and of some vertices in the de Bruijn trees. 

\begin{thm}\hspace*{-0.60em}\textbf{.}\label{thm5} 
If $\emph{Card}({\cal A})=q$ and $q^k+k\le N \le q^{k+1}+k$ then $M(N)$ is equal to
the number of different paths of length $N-k-1$ in the de Bruijn graph
$B(q,k+1)$.
\end{thm}

\medskip
\noindent\textbf{Proof.} 
From Propositions 1 and 2 it follows that the number $M(N)$ of the words
of length $N$ with global maximal complexity is given by the number of words 
$w\in {\cal A}^N$ with $f_w(k+1)=N-k$. It means that these words contain 
$N-k$ subwords of length $k+1$, all of them distinct. To enumerate all of
them we start successively with each word of $k+1$ letters (hence with each
vertex in $B(q,k+1$)) and we add at each step, in turn, one of the symbols
from ${\cal A}$ which does not duplicate a word of length $k+1$ which has 
already appeared.  Of course, not all of the trials will finish in a word 
of length $N$, but those which do this,  are precisely paths in $B(q,k+1)$
starting with each vertex in turn and having the length $N-k-1$. Hence to 
each word of length $N$ with $f_w(k+1)=N-k$ we can associate a path and 
only one of length $N-k-1$ starting from the vertex given by the first
$k+1$ letters of the initial word; conversely, any path of length
$N-k-1$ will provide a word  $w$ of length $N$ which contains $N-k$ 
distinct subwords of length $k+1$.

\medskip
\begin{rem}\hspace*{-0.60em}\textbf{.} 
\emph{
The number of words of length $N$ having global maximal complexity  can be also expressed by means of certain vertices in the de 
Bruijn trees. $M(N)$ is equal to the number of vertices at the level 
$N-k-1$ in the set $\{ T(q,w) \;|\;w\in {\cal A}^{k+1} \}$ of the de 
Bruijn trees. (The level of the root is considered to be 0, its descendants 
are on level 1 etc.) }

\emph{
The other four trees corresponding to the de Bruijn graph $B(2,3)$ are mirror images of those in Fig. 
\ref{fig3a}--\ref{fig3d}; we obtain, for example,
$M(6)$ by doubling the number of vertices at level 3 in Fig. 
\ref{fig3a}--\ref{fig3d},
 i. e. $M(6)=2\cdot 18=36.$ Similarly $M(7)= 2\cdot 21 =42$ is obtained by doubling the
number of vertices at level 4, and so on up to $M(10)=2\cdot 8=16 $ (using the vertices at level 7).
These results are in accordance with those given in Table 1 obtained by counting all possible
words with maximal complexity.}
\end{rem}

\medskip
A formula for the number $M(N)$ of the words whose maximal complexity is equal to the global maximal complexity $K(N)$ can be given for the special case of de Bruijn words.

\medskip
\begin{thm}\hspace*{-0.60em}\textbf{.} 
If $N=2^k+k-1$ then $M(N)= 2^{2^{k-1}}$.
\end{thm}

\medskip
\noindent\textbf{Proof.}
The number of distinct Hamiltonian cycles in the de Bruijn graph $B(2,k)$ 
is equal to $\displaystyle{2^{2^{k-1}-k}}$ \cite{bond}. 
With each vertex of a Hamiltonian cycle a de Bruijn word (containing all the
factors of length $k$) begins, which has 
maximal complexity, so $M(N)=\displaystyle{2^k\cdot 2^{2^{k-1}-k}}$, which 
proves the proposition.  (In \cite{bruijn} the number of circular de Bruijn 
words is found, which corresponds to the number of Hamiltonian cycles in de 
Bruijn graphs).

\medskip
A generalization for $q\ge 2$ can be proved in a similar way using the results 
in \cite{aardenne}. 

\begin{thm}\hspace*{-0.60em}\textbf{.} 
If $N=q^k+k-1$ then $M(N)= (q!)^{q^{k-1}}$.
\end{thm}

\medskip
In Proposition 1, respectively Proposition 2, we have determined for each natural number $N$ the value of the global maximal complexity $K(N)$, respectively the set of values $i$ for which there exists a word of length $N$ with $K(N)$ subwords of length $i$. To obtain a general formula for $M(N)$ for each natural number $N$ is still an open problem.


\begin{thebibliography}{9}

\bibitem{aardenne} T. van Aardenne-Ehrenfest, N. G. de Bruijn, 
\textit{Circuits and trees in oriented linear graphs}, Simon Stevin
 28 (1951), 203--217.

\bibitem{bond} J. Bond, A. Iv\'anyi,  \textit{Modelling of interconnection
networks using de Bruijn graphs}, Third Conference of Program Designer, 
Ed. A. Iv\'anyi, Budapest, 1987,  75--87.

\bibitem{bruijn} N. G. de Bruijn, \textit{A combinatorial problem}, 
Nederl. Akad. Wetensch. Proc. 49 (1946), 758--764.

\bibitem{hei} M. Heinz, \textit{Zur Teilwortcomplexit\"at f\"ur 
W\"orter und Folgen \"uber einem endlichen Alphabet}, 
EIK 13 (1977), 27-38.

\bibitem{lothaire} M. Lothaire, \textit{Combinatorics on words}, 
Addison-Wesley, Reading, MA, 1983.

\bibitem{aldo} A. de Luca, \textit{On the combinatorics of finite words}, 
Theor. Comput. Sci. 218 (1999), 13--39.

\bibitem{martin} M. H. Martin, \textit{A problem in arrangements}, 
Bull. A. M. S. 40 (1934), 859--864.

\end{thebibliography}
\end{document}